# Simulation of the flyby anomaly by means of an empirical asymmetric gravitational field with definite spatial orientation.

H. J. Busack

Wulfsdorfer Weg 89, 23560 Lübeck, Germany
27. November 2007

## Abstract

All anomalous velocity increases until now observed during the Earth flybys of the spacecrafts Galileo, NEAR, Rosetta, Cassini and Messenger have been correctly calculated by computer simulation using an asymmetric field term in addition to the Newtonian gravitational field. The specific characteristic of this term is the lack of coupling to the rotation of the Earth or to the direction of other gravitational sources such as the Sun or Moon. Instead, the asymmetry is oriented in the direction of the Earth's motion within an assumed unique reference frame. With this assumption, the simulation results of the Earth flybys Galileo1, NEAR, Rosetta1 and Cassini hit the observed nominal values, while for the flybys Galileo2 and Messenger, which for different reasons are measured with uncertain anomaly values, the simulated anomalies are within plausible ranges. Furthermore, the shape of the simulated anomaly curve is in qualitative agreement with the measured Doppler residuals immediately following the perigee of the first Earth flyby of Galileo. Based on the simulation, an estimation is made for possible anomalies of the recently carried out flybys of Rosetta at Mars on 25.02.07 and at the Earth on 13.11.07, and for the forthcoming Earth flyby on 13.11.09. It is discussed, why a so modelled gravitational field has not been discovered until now by analysis of the orbits of Earth satellites, and what consequences are to be considered with respect to General Relativity.

## 1 Introduction

During several Earth flybys carried out since 1990, some spacecrafts have experienced an unexpected and until now unexplained anomalous velocity increase by a few mm/s. This phenomenon is called the *flyby anomaly* and looks like the effect of an instantaneous acceleration of the spacecraft at the time of closest approach to Earth. The measured velocity differences could be approximately reproduced by estimation of a sudden change of the velocity vector at perigee [2].
Another approximate simulation was possible by means of estimation of the gravitational harmonic coefficients J2, C21, S21, C22 and S22 of the Earth's gravity field. However, the estimated values for these coefficients were by far unreasonable, and moreover different for the Galileo1 and the NEAR flyby [2].
Many efforts have been made to find a reason for this acceleration, but none of them were able to give an explanation based on known physics.
Likewise without success were attempts to modify the law of gravitation in order to get better agreement with the observations. A good summary of these efforts is given by [2, 4, 5]. Additionally, a comprehensive overview of gravity tests is provided by [6].
Further attempts were made to find at least phenomenological patterns in the flyby data, i.e., relationships between eccentricity, perigee altitude and perturbations by the Sun, Moon or planets [3, 4]. Due to the poor database of only a few flybys, these attempts were without success, as well.
Hence it is yet unclear, whether an unconsidered interaction based on known physics is responsible for this mysterious effect, or if the known physical laws, especially General Relativity, have to be modified. After more than 15 years of careful analyses it seems unlikely to find an explanation based on the standard physical laws.
Therefore, in this investigation another attempt was made to find an addendum to Newton's law of gravitation, which could simulate all observed anomalies, at first without any regards to a consistent theory.
To this end three basic assumptions for the structure of the additional gravitational field term were made:

1. No spherical symmetry to the center of the Earth
2. No fixed orientation of the asymmetry to the surface coordinate system of the Earth
3. Short interaction distance

The reason for item 1 seems to be obvious, since in the case of spherical symmetry and short interaction distance of the additional field one would expect no difference to the velocity magnitude of a Newtonian trajectory for long and equal geocentric distances because the trajectory would be symmetric, too. This has been proved by numerical integration of trajectories with Newtonian and symmetric non-Newtonian gravity fields.

Item 2 is necessary, because otherwise the asymmetry would easily be found by analysis of the orbits of Earth satellites used for determination of the Earth's gravity field. This point will be discussed later in more detail.

Item 3 is justified by the lack of an unexplained acceleration in the motions of geostationary satellites, and by the nearly instantaneous onset of the acceleration near the point of closest approach (CA) of a flyby trajectory.

Based on these assumptions, a more specific mathematical formulation for a numerical simulation was found. This formula was implemented in a computer program for simulating the above-mentioned flybys. The free parameters of the equation were varied in order to achieve compliance with the measured anomaly values of all flybys. Clearly such simulation is far away from a consistent theory, but, if successful, can potentially show the direction of a next step.

## 2   Justification of the empirical field term

Of course, the necessary properties of the hypothetical field term stated in the introduction are not sufficient. Below is set out, which reasons have led to the concrete formulation of the gravitational model.

The rapid decrease of the additional field term with greater distance from the surface of the Earth, the asymmetry and the lack of coupling to the Earth's rotation suggests that the source of the anomaly is not outside the Earth nor related to the mass distribution of the Earth, but is an unknown intrinsic mass property, in combination with a certain spatial direction. As a plausible direction, the apex of the Earth's motion with respect to the reference frame of the cosmic microwave background (CMB) was investigated. The CMB-apex of the Earth can be derived from the CMB-apex of the Sun, the Sun velocity within the CMB reference frame and the velocity vector of the Earth at the Sun orbit for a given time. The coordinates of the CMB-apex of the Sun are:

RA:   11h 12m
DEC:  -7° 13'

The magnitude of the Sun velocity in this direction is 369km/s [1].

The first working hypothesis was, that the motion of the Earth's mass (or any other mass) with respect to the CMB reference frame leads to an additional gravitational field term, proportional to the mass, rapidly decreasing with greater distance to the mass, and asymmetrically deformed in the direction of motion, whereas the asymmetry and possibly the amplitude of the additional term are increasing with the velocity magnitude relative to the CMB rest frame. Since the velocity of the Sun has to be considered as constant in this reference frame, the direction of the Earth's motion is nearly spatially-fixed due to the much lower orbit velocity of the Earth.
Indeed, the assumption of a measurable action of a straight, constant motion against a unique reference frame (following, referred to as gravitational rest frame) is contrary to the Principle of Relativity, one of the main principles both of Special and General Relativity. This point will be discussed later in more detail.

As the most simple but natural model of a rapidly decreasing function the exponential function was chosen. The asymmetry was implemented in two different ways. Firstly, as asymmetry of the slope constant of the exponential function, secondly, as asymmetry of the amplitude, both with harmonic characteristic of first order. This leads to the following equation for the hypothetical gravitational field $\vec{g}(\vec{r})$ in the exterior space of a spherically symmetric mass distribution as a starting point of the investigations:

$$\vec{g}(\vec{r}) = -\frac{G \cdot M \cdot \vec{r}}{r^3}\left[1 + A \cdot F_A \cdot \exp\left(-\frac{r-R}{B - C\dfrac{\vec{r} \cdot \vec{v}}{r \cdot v_{Sun}}}\right)\right] \tag{1}$$

with $r = |\vec{r}| \geq R$ and

$G$: gravitation constant
$M$: field mass
$R$: radius of the field mass body
$\vec{r}$: position vector of the test mass with origin at the center of the field mass body
$\vec{v}$: velocity vector of the field mass center in the gravitational rest frame
$v_{Sun}$: magnitude of the Sun velocity in the gravitational rest frame
$F_A$: amplitude factor
$A, B, C$: arbitrary constants

The amplitude factor $F_A$ rules the amplitude behaviour of equation (1). The following cases with different dependency on the velocity of the mass against the gravitational rest frame were investigated:

$$F_A = 1 \tag{2}$$

This is the case with no influence on the amplitude of the additional field term.

$$F_A = \frac{v}{v_{Sun}} \tag{3}$$

Here the amplitude is proportional to the velocity of the mass.

$$F_A = \frac{\vec{r} \cdot \vec{v}}{r \cdot v_{Sun}} \tag{4}$$

Sign and magnitude of the amplitude depends on the velocity and position vectors.

$$F_A = \frac{v}{v_{Sun}} + D\frac{\vec{r} \cdot \vec{v}}{r \cdot v_{Sun}} \tag{5}$$

The asymmetry of the amplitude and the amplitude depend on the velocity, where $D$ is an arbitrary constant, at first set to 1 (limit case without sign change).

So there were four different classes of equation (1) for the first investigations. What class could be better adapted to the measured anomalies had to be decided by the simulation.

To do so, for each of the four function classes the arbitrary parameters *A*, *B* and *C* were varied, in order to achieve agreement of the simulation with the anomalies of the three best observed flybys. These are the Earth flybys Galileo1, NEAR and Rosetta1, for which close tolerances of the measured anomalies are stated [4, 5].Further degrees of freedom for the simulation are delivered by magnitude and direction of the velocity in case that the hypothetical gravitational rest frame does not coincide with the CMB reference frame. In fact, it has turned out, that this is necessary for a successful simulation using the mentioned classes of equation (1).

## 3    Realization of the simulation

The simulation was written as a computer program by means of a high-level programming system of the language family PASCAL. The trajectory data of each flyby were taken from the HORIZONS web interface of JPL [7], the coordinates at closest approach (CA) were used to get the velocity vector of the spacecraft and the time of CA.
With the aid of CA time, the orbit position of the Earth, and hence its velocity vector within the hypothetical gravitational rest frame, was determined.
With position and velocity of the spacecraft at CA the trajectory was calculated back with time steps of 10ms until 5 hours before CA, using accelerations according to Newton's law of gravitation for a spherical symmetric model of the Earth.
At this point, the velocity vector components were inverted, establishing in this way the starting condition for the simulation. With this starting condition, the trajectory of the spacecraft was calculated both with accelerations after Newton's law and with the accelerations after equation (1), again with 10ms steps until 5 hours after CA. The data of the Newtonian trajectory at CA were compared with the original data in order to get a measure for the quality of the simulation.
The restriction to a spherical symmetric model of the Earth without any multipole coefficients or perturbations by other gravity sources was regarded as sufficient for the intended simulation purpose, because these effects are nearly the same for both trajectories and are mostly cancelled out by calculating the difference of the velocities of both trajectories in order to obtain the flyby anomaly. This assumption has been proved by calculating the flyby anomalies both with and without the presence of an additional gravity acceleration of 0.013m/s² in a certain direction in the projection area of the Earth model. This acceleration value is the maximum acceleration seen by the NEAR spacecraft due to the quadrupole moment of the Earth [2]. The differences for the simulated flyby anomalies were in all cases less than 0.03mm/s, so the assumption is justified.
To find a reasonable measure for the velocity difference of both trajectories, different possibilities were investigated.
Firstly, the difference of the velocity magnitudes of both trajectories for equal time, secondly, the difference for equal geocentric distances and thirdly, the difference for equal track position.
The first possibility has no constant limit value for great distances; the second possibility gives no reasonable values in the vicinity of CA. As a good compromise between both possibilities, the difference for equal track position was chosen. This criterion was implemented by normal projection of the respective Newtonian track point onto the hypothetical track.
As numerical value for the flyby anomaly, the difference at 5 hours after CA was taken. This value has been proved as good representant for the value at 'infinity' in most cases. For anomaly curves with excessive overshoot after CA, the anomaly value at 100 hours after CA was used.
For reference purposes, a download of the executable program file flyby_anomaly.exe is provided by [8]. The source code can be requested from the author.

# 4 Results of the simulation

## 4.1 Preliminary investigations

The first attempt was, to reach the above stated three anomaly values within their tolerances by variation of the parameters *A*, *B* and *C* of equation (1) without regard to the other flyby anomalies or to the shape of the anomaly curve.
With the assumption of the CMB reference frame as the hypothetical gravitational rest frame, no compliance with the measured anomaly values could be found using all four classes of equation (1).

Therefore the velocity vector within the gravitational rest frame also was varied, leading to grossly increased calculation effort. For such from the CMB reference frame different gravitational rest frame, compliance was reached with three of the four field term classes. Regarding the shape of the curve and the anomaly values of the less precisely observed flybys, there were larger differences, allowing a first evaluation.

Following, the tabular results of the three models are given (table 1 and table 2):

**Table 1 :**
Simulation parameters derived by preliminary investigations with different function classes

| Parameter | Equation (2) | Equation (3) | Equation (5) |
|---|---|---|---|
| *A* | 0.0002186 | 0.0002609 | 0.0019200 |
| *B* / m | 293000 | 284000 | 290000 |
| *C* / m | 234224 | 228000 | 238000 |
| $V_S$ / m/s | 295000 | 315000 | 295000 |
| *RA* / h | 16.33 | 16.48 | 12.78 |
| *DEC* / ° | -40 | -38.83 | 14.83 |

**Table 2 :**
Results of the preliminary investigation with different function classes.
Measured values after [4, 5]

| Earth flyby | Flyby anomaly / mm/s | | | | | comment |
|---|---|---|---|---|---|---|
| | Simulated | | | Measured | Tolerance | |
| | equ.(2) | equ.(3) | equ.(5) | | | |
| Galileo_1 | 3.92 | 3.92 | 3.92 | 3.92 | 0.08 | 08.12.90 |
| NEAR | 13.46 | 13.46 | 13.72 | 13.46 | 0.13 | 23.01.98 |
| Rosetta_1 | 1.82 | 1.82 | 1.79 | 1.82 | 0.05 | 04.03.05 |
| Cassini | 0.00 | 0.00 | -2.37 | 0.11 | -- | 18.08.99 |
| Galileo_2 | 4.02 | 3.83 | -21.04 | -- | -- | atmospheric drag |
| Messenger | 0.84 | 0.84 | 1.20 | ~0 | -- | no reliable data |
| Rosetta_Mars | 9.52 | 11.07 | -28.31 | -- | -- | 25.02.07 |
| Rosetta_2 | 0.00 | 0.00 | 0.00 | -- | -- | 13.11.07 |
| Rosetta_3 | 0.00 | 0.00 | 0.14 | -- | -- | 13.11.09 |

## 4.2 Further improvement

Galileo1 is the only well-observed Earth flyby with 2-way-Doppler data close to the time of CA. In [2] the data are graphically presented. They show a distinct overshoot of approximately double the final anomaly value. A similar overshoot showed all simulated anomaly curves. Unfortunately, the temporal resolution of this graph is poor, so no meaningful comparison with the simulation could be done without additional information. Furthermore, there are unclarities regarding the temporal mapping of the respective DSN stations, and discrepancies of data, due to low elevation. For example, Antreasian and Guinn [2] stated, that Goldstone obtained 2-way Doppler data at 14 minutes after encounter, while Canberra acquired lock "a few minutes after the short 30 minute Goldstone track". This is an ambiguous statement; possible meanings are a few minutes after the beginning, or a few minutes after the end of the Goldstone track. If a simultaneous 2-way Doppler contact with two stations was technically possible, the first alternative seems to be more plausible, because at the latest 30 min after CA Canberra was in much better position for observation, regarding the elevation. After that, one has to consider, that the Canberra data track may have begun either approximately 25 minutes or 47 to 55 minutes after CA. In any case, the data of Goldstone are not compatible within this time span with the Canberra data and were discarded due to low elevation for the delta-v estimation of the Galileo1 flyby [2]. Anderson et al. [5] also stated that the time history of the velocity increase of Galileo1 was not obtainable.

The simulation with equation (5) and D=1 shows for Cassini, Galileo2 and Messenger anomalies, which are probably not consistent with observation. Furthermore, the curve of Galileo1 shows a very distinctive overshoot that exceeds the observed values even in the case that the Canberra data starts approximately 55 minutes after CA. For a better adaption to the measured data, in equation (5) the consideration of a further arbitrary constant (D) would be necessary, further increasing the calculation effort, without knowing, whether the late Canberra track is the right one.

With equation (4) no compliance with observation was investigated.

The simulation results with equation (2) and equation (3) show very similar curves in qualitative agreement with the Galileo1 curve and similar anomalies with acceptable values for the other flybys.

For these reasons, the simplest case with $F_A$=1 (equation (2)) was chosen for further optimization with respect to the curve and to the anomalies of the remaining flybys.

In order to make quantitative comparisons in spite of the above-mentioned unclarities, the Doppler residuals of Galileo1 were used at first only for additional information. Possibly better data were expected from the acceleration curves of the NEAR flyby, obtained from a hypothetical Earth gravity field estimation, which could reduce the Doppler residuals to an acceptable value [2]. These data are available with good temporal resolution.
The gravitational harmonic coefficients J2, C21, S21, C22 and S22 of the Earth gravity field could be estimated by [2] in a way, that the discontinuities in the Doppler residuals of the Galileo1 and the NEAR flyby are removed. The coefficient values were by magnitudes out of the acceptable tolerance limit, therefore being unreasonable and furthermore different for both flybys. But with these coefficients, it was possible to calculate the acceleration curves for the hypothetical gravity fields, and hence the differences to the nominal model. So one can get acceleration data, which simulate approximately the measured anomalies, and therefore are possibly suitable for a comparison, even in the case, that the original data are not available.

The acceleration values for the NEAR flyby are given by [2] as components along track, in radial and transverse direction and as magnitude. The components are shown here in figure 1.

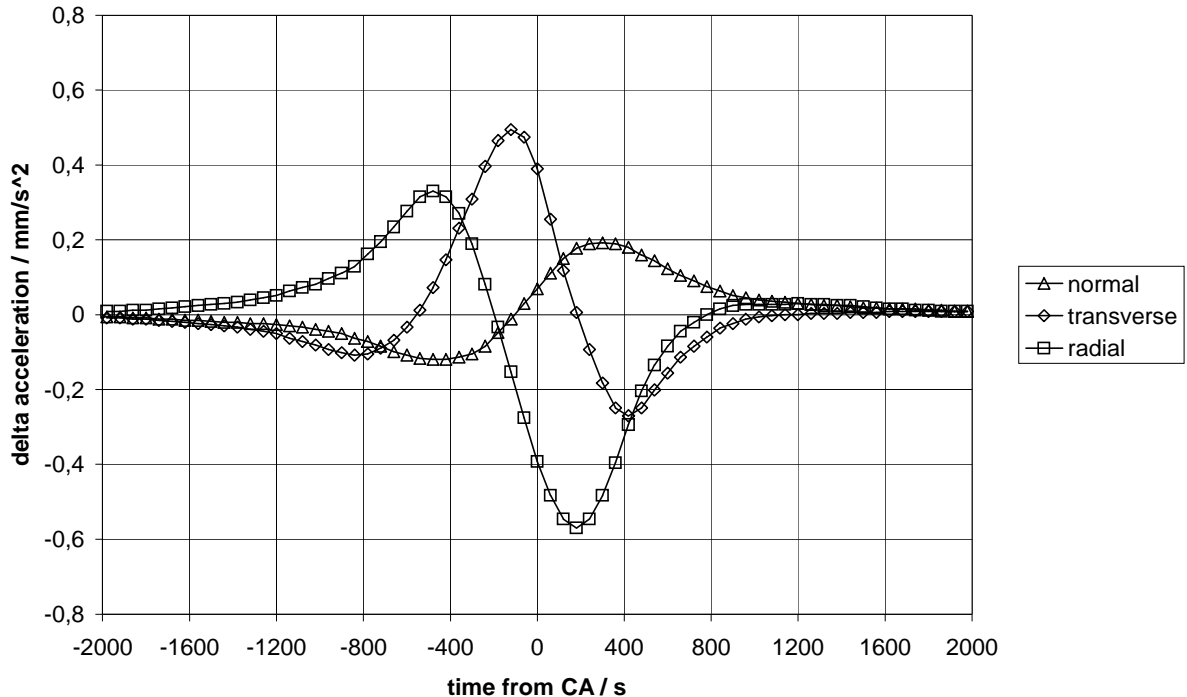

**Figure 1:** Differences between the hypothetical and the nominal JGM-3 gravity accelerations for the NEAR Earth flyby – after [2]

For the presented simulation these acceleration components were calculated as well, and compared with figure 1.
The result was, that with the parameters of the preliminary survey only the transverse component showed a similar form, but with less amplitude.
The normal component was only roughly similar.
The largest differences showed the radial component. The amplitude was less than stated by [2], as for the other components, but the curve has values of but one sign, so the net acceleration was far higher.

Because of these discrepancies, the next attempt was to find simulation parameters with not only correct anomaly values, but with better matching acceleration curves for the NEAR flyby. This attempt was without satisfying success; in no case were nearly equal accelerations found.
With hindsight it can be stated, that a good agreement could not be expected, because the former estimation was approximate and with unreasonable values.
So this way for getting quantitative comparison for the anomaly curve was unusable, as well. The proof for obtaining better matching curves with the proposed field term must be left to further investigations.

But, as an extra benefit of these efforts, parameters were found, which additionally hit the nominal value of the Cassini flyby, further giving a smaller value for Messenger and simultaneously a good matched curve for Galileo1 for the case (if possible and real) of an early Canberra 2-way Doppler track starting approximately 25 to 30 minutes after CA.

Furthermore, parameter sets were found for curves that matched well a late Canberra track starting at 47 minutes after CA. The characteristic of these curves is a higher overshoot of the anomaly values immediately after CA over the respective limit values at "infinity".

Two of these parameter sets and the simulation results are presented in table 3 and table 4.

**Table 3 :** Final parameter sets optimized with equation (2), "low" and "high" overshoot

| overshoot | Parameter | | | | | |
|---|---|---|---|---|---|---|
| | A | B / m | C / m | $V_S$ / m/s | RA / h | DEC / ° |
| "low" | 2.453e-4 | 394000 | 136000 | 360000 | 17.78 | -37.50 |
| "high" | 2.306e-4 | 464000 | 105000 | 360000 | 17.70 | -61,00 |

**Table 4 :**
Simulation results of the final optimization with equation (2). Measured values after [4, 5]

| flyby | flyby anomaly / mm/s | | | | comment |
|---|---|---|---|---|---|
| | Sim. "low" | Sim. "high" | meas. | tol. | |
| Galileo_1 | 3.92 | 3.92 | 3.92 | 0.08 | 08.12.90 |
| NEAR | 13.46 | 13.88 | 13.46 | 0.13 | 23.01.98 |
| Rosetta_1 | 1.82 | 1.89 | 1.82 | 0.05 | 04.03.05 |
| Cassini | 0.11 | 0.22 | 0.11 | -- | 18.08.99 , thruster activities |
| Galileo_2 | 5.54 | 5.50 | -- | -- | 08.12.92 , atmospheric drag |
| Messenger | 0.69 | 0.80 | ~0 | -- | 02.08.05 , no reliable data |
| Rosetta_Mars | 6.72 | 4.20 | -- | -- | 25.02.07 , yet under evaluation |
| Rosetta_2 | 0.00 | 0.00 | -- | -- | 13.11.07 |
| Rosetta_3 | 0.00 | -0.03 | -- | -- | 13.11.09 |

As one can see, the congruence of the "low" simulation with the measured data is excellent, and very encouraging for further investigations with potentially better data or with additional data. The data of simulation "high" differ somewhat from the nominal values. This is due to the steeper curves, not reaching the limit value within 5h after CA. In this case, the parameters were chosen in such a way, that the simulated values of the first three anomalies were well within the respective tolerances at 100h after CA.

Because without better or new data no possible improvement could be seen in this investigation, the simulation has been stopped at this point. For the same reason, no systematic investigations were carried out to determine, how close the computed 5h anomaly values were related to the limit values at infinity. Sporadic simulation tests up to 100h after CA show, that for "low" parameters the error at 5h after CA in all cases was less than +/- 0.08mm/s.

Likewise no further attempts were made to increase the accuracy of the simulation by considering the multipole coefficients of the gravity field of the Earth or perturbances by other gravity sources like Sun or Moon. As mentioned above, for flybys with low perigee altitude, the accuracy could be gained by some 0.03mm/s.

Two further interesting aspects should be pointed out. Parameter $V_S$ is of little influence to the results and was held constant throughout the final optimizing process. Firstly, this implies that the remarkable compliance of 6 anomaly values and additionally of one curve shape was practically reached by variation of only 5 arbitrary parameters. This seems to be a clear indication of an underlying physical reality in contrast to an only numerical effect. Secondly, the reason for the asymmetry of the gravitational field is not restricted to the proposed absolute motion, but could also be caused by a non-isotropic property of spacetime.

With the final "low" parameter set, the following acceleration curves were obtained for the NEAR-flyby (Figure 2):

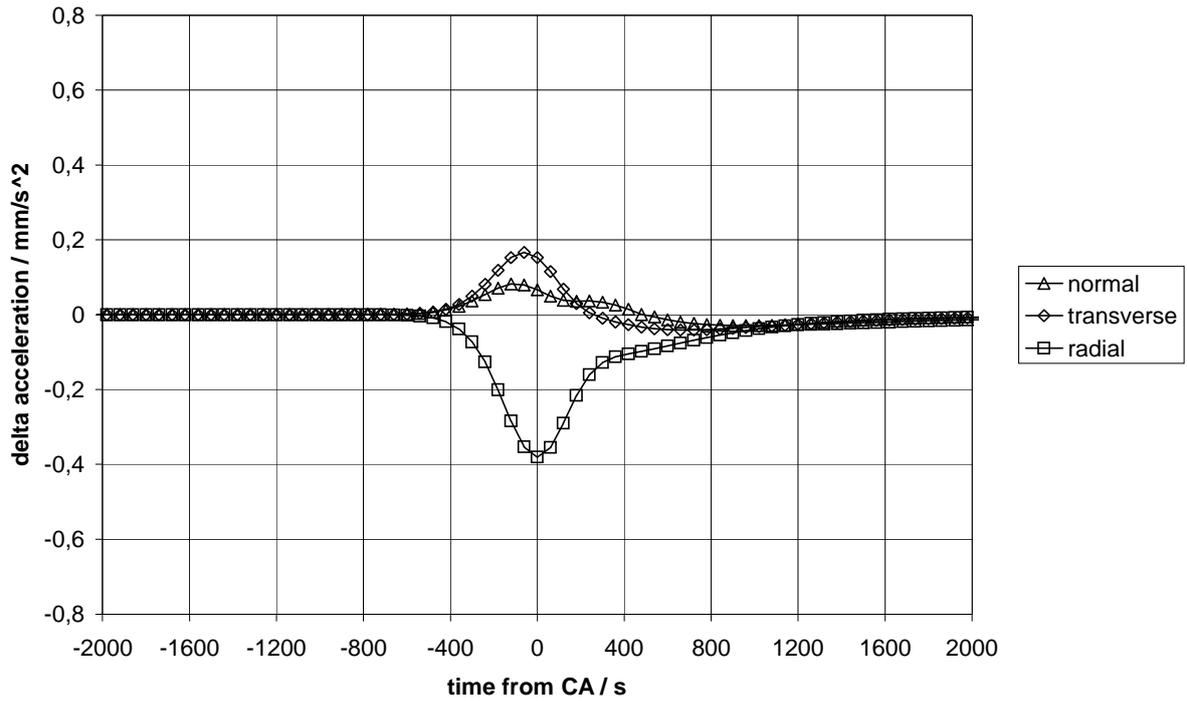

**Figure 2:** Differences between the proposed non-Newtonian (equation (2), optimized) and the Newtonian gravitational accelerations for the NEAR Earth flyby

For the Galileo1 flyby the following anomaly curves were obtained with the final parameter sets from table 3 (Figure 3):

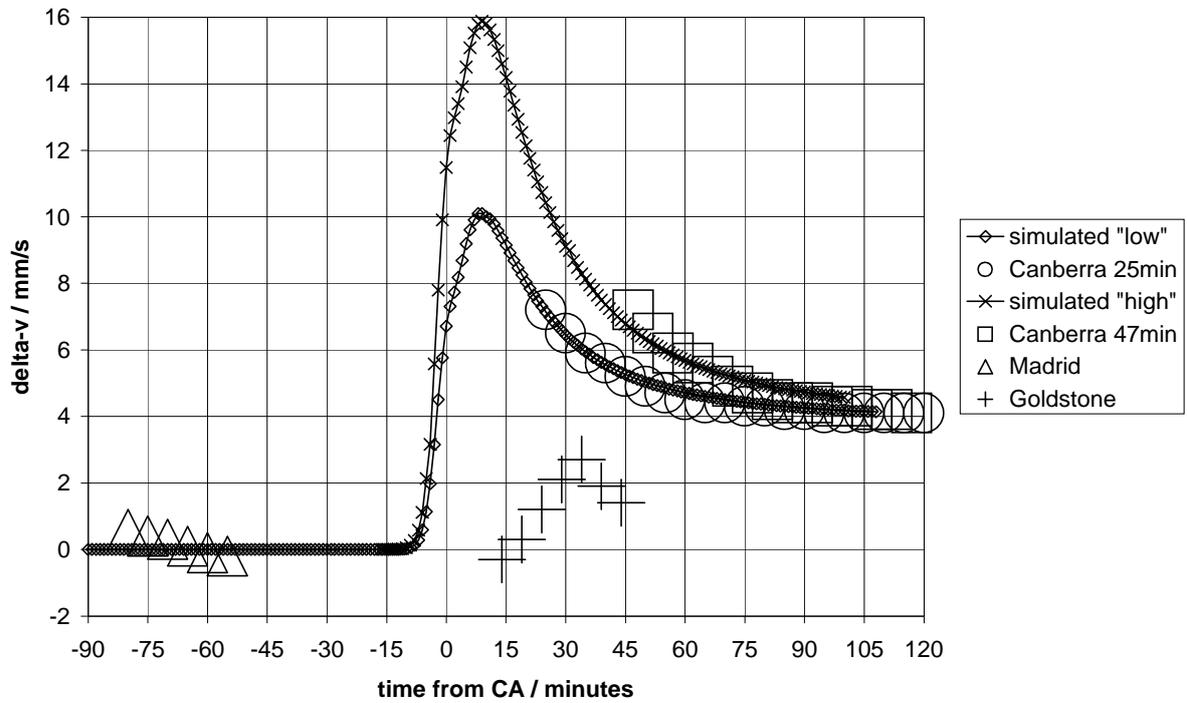

**Figure 3:** Simulated delta-v anomaly at Galileo1 Earth flyby, optimized parameters with equation (2). Measured data estimated after [2]

The measured values in Figure 3 are estimated from the Doppler residual graph in [2], while the time scale is reconstructed from the associated text. Since it is unclear, whether the early or the late Canberra track is the right one, both possibilities are shown. Because of the great uncertainty of the estimated data, the congruence of the curves cannot be considered as quantitative but rather as qualitative. For better evaluation, access to the original residual data would be neccesary.

# 5  Why has the anomalous gravitational field been unexplored until now?

If the proposed gravitational field or a similar one is actually real, one has to answer, why it has not been discovered by the analysis of satellite orbits until now. Several reasons are supposable, discussed in three categories following:

## 5.1  coverage by tidal forces at medium term

The proposed gravitational field is asymmetric with nearly spatially-fixed orientation. At medium term (i.e., for several weeks) therefore the orientation to the tidal force of the Sun stays nearly constant. Within such a time interval the effect would be misinterpreted as a deformation of the tidal coefficients. The change over longer periods could be assigned to annual effects.

## 5.2  Distinctive dependency on altitude and orbit orientation

Due to the exponential function, the altitude dependency of the additional field term is very distinct, and because of the asymmetry, moreover, different for the diverse orbit orientations of satellites. Hence the contributions of satellites to the determination of the Earth's gravity field are different with respect to known physics, therefore regarded as inconsistent and mostly are cancelled out as noise. As an example for the altitude and direction dependency, the results for the "low" parameter set of table 3 are given in table 5.

**Table 5 :**
Values of the additional field term related to the Newtonian field with final "low" parameters

| $\theta$ / ° | h / km | | | | | |
|---|---|---|---|---|---|---|
| | 0 | 1000 | 2000 | 3000 | 4000 | 5000 |
| 0 | 2.45e-04 | 5.09e-06 | 1.05e-07 | 2.19e-09 | 4.53e-11 | 9.40e-13 |
| 45 | 2.45e-04 | 8.54e-06 | 2.97e-07 | 1.04e-08 | 3.61e-10 | 1.26e-11 |
| 90 | 2.45e-04 | 1.94e-05 | 1.53e-06 | 1.21e-07 | 9.56e-09 | 7.56e-10 |
| 135 | 2.45e-04 | 3.19e-05 | 4.15e-06 | 5.39e-07 | 7.01e-08 | 9.11e-09 |
| 180 | 2.45e-04 | 3.72e-05 | 5.63e-06 | 8.54e-07 | 1.29e-07 | 1.96e-08 |

The values are given for different altitudes h and angles $\theta$ to the center of Earth. Zero point of the angle $\theta$ is in the direction of the apex in the assumed gravitational rest frame.

Acceleration differences of the order $10^{-7}$ of the Earth's gravity acceleration are hardly detectable. This corresponds to values of approximately 1e-7 in table 5. Therefore at perigee altitudes considerably above 4000km, in no direction regarding the apex are differences detectable, while at altitudes of about 2000km the anomalies vary between not detectable in the apex direction and detectable in the opposite direction. At altitudes below 300km the anomaly would be easily detectable in any direction, but the perturbations by atmospheric drag are not so well modelled that the anomaly values could be derived with certainty.

## 5.3 Filter effect of orbit determination or analysis software

Even without deeper knowledge of the software used for determination of the Earth's gravitational field, one can assume, that diverse filter algorithms are used to identify the contributions of different gravitation sources. That would have the consequence, that all sources without a matching characteristic are suppressed as noise effects. Because obviously until now no spatially-fixed gravitation source has been taken into account, probably no suitable filter exists or was used, to find such a component.

## 6 Consequences to General Relativity

As stated in the introduction, the assumption of a gravitional field term caused by an 'absolute' motion of the mass body against a unique reference frame, violates the Principle of Relativity. Because both Special and General Relativity are based on this principle, the proposed field cannot be derived from current theory. This would even hold true, if the asymmetry would not be caused by the motion of the mass, but rather by a non-isotropic property of spacetime in combination with the mass.

In case, this or a similar gravitational field turns out to be real, attempts have to be made to develop a new theory, which correctly describes the new phenomena, while covering the very well confirmed General Relativity as a limit case. What kind of limit case this would be depends on the cause for the asymmetry of the gravitational field. The motion against a gravitational rest frame, considered here, would require evanescent velocity within this frame or at least sufficient distance to mass concentrations and would probably lead to a theory not including the Principle of Relativity in the present form.

What approach to a new theory can be derived from the structure of the additional field term? No consistent explanation could be found within this investigation for the rapid decrease of the filed term with the distance from the surface of Earth (boundary of the mass concentration), but at least some hints. The exponential decay and the asymmetry caused by a motion against a unique reference frame leads one to think of an unsteady transport phenomenon within this reference frame. To prove this assumption, numerical calculations have been made using a parabolic partial differential equation with different source fields moving with constant velocity in a 'transport' reference frame. No compliance between the calculated transport field in the vicinity of the source field and the gravitational field after equation (1) was found for a spherical source in three spatial dimensions. A circular source in three spatial dimensions was insufficient, as well. Similar but not exactly matching transport fields were found for the case of an annular source in three spatial dimensions. Here the calculated field in the plane of the annular source showed exponential decay with distance from the annular source, as to be expected from equation (1). But the asymmetry of the field has a harmonic characteristic of higher order, in contrast to equation (1). One possible approach to this case would be a hypothetical source of the transport phenomenon at a mass density step. This source would be located in the surface of a mass body and could possibly act as a source of an unsteady transport phenomenon with nearly matching geometry in the case of motion against a hypothetical transport reference frame of four spatial dimensions, analogously to the annular source in three dimensions. This idea was not carried out in detail, due to the lack of foundation in a consistent theory, and because the simulation would take an enormous amount of calculation time. So no well-founded explanation can be given by this investigation.

In case the additional field term could actually be described as a transport phenomenon, this would imply that a part of the gravitational field could exist without coupling to a mass, but with coupling to the transport reference frame (i.e., the gravitational rest frame). In the case of flyby anomaly this part is obviously nearly negligible small. But if decoupling is possible in principle, one can suppose, that under other circumstances, i.e., in cosmic dimensions, significant parts can decouple. Such decoupled fields would probably take part on the cosmic expansion, in contrast to gravity bound matter condensations. This in turn leads one to think of Dark Matter, for which clues are present, that coupling to baryonic matter is not seen in all cases. If this holds true, the validity of General Relativity would not only be violated for the vicinity of mass concentrations, but also in cosmic dimensions, dependent on the previous history. The latter conclusions are highly speculative, but show a possible potential for the suggested approach. The embedding of such an approach into a theory of relativity is most probably a big challenge.

# 7 Predictions and testing feasibilities

## 7.1 Anomalies at further flybys

The preliminary investigation with different variants of the additional field term and also the optimizing phase with equation (2) has led to a plurality of parameter sets for which the anomalies of the best observed flybys are fully, and the anomalies of the rest are more or less correctly, simulated. These parameter sets were used also for the Mars flyby, and for the forthcoming Earth flybys of Rosetta. Whether for a Mars flyby the same parameters are to be used as for an Earth flyby, is not clear *a priori*. For example, the parameters could depend on mass, but also on the diameter of the mass body. In spite of this uncertainty the Earth parameter sets were tentatively used for Mars. While the so simulated anomalies for the Mars flyby showed a considerable variation, even with negative values, the simulations for the Rosetta Earth flybys have led without any exception to nearly unmeasurably low anomaly values. Because of the above-mentioned incomplete database, no systematic error analysis could be done. Instead, the following predictions for the anomalies of the yet unevaluated Mars flyby and for the forthcoming Earth flybys of Rosetta are done by estimation based on the calculation experience acquired during the simulation process:

Mars  25.02.07:      (   5 +/-    3) mm/s
Earth 13.11.07:      (0.00 +/- 0.01) mm/s
Earth 13.11.09:      (0.00 +/- 0.04) mm/s

As mentioned above, the simulated anomalies are valid for the Mars flyby only in the case that the simulation with equation (2) is feasible, and but here in the case, that the Earth parameters could be used for Mars. If this is not the case, significant differences are to be expected, but the absence of a measurable anomaly would be unlikely for a well-known Martian gravity field, because the diameters of Mars and Earth are of the same order and the reality of the assumed mass proportionality can be considered as plausible. The gravity field of Mars is currently being precisely explored by Mars Reconnaissance Orbiter. It is to hope, that the precision of the results will be sufficient for detecting an anomaly at the Rosetta flyby.
If an anomaly at the Rosetta Mars flyby is actually seen, the reality of a phenomenon, unexplainable with current physics, has to be accepted. Especially the forthcoming Earth flybys of Rosetta would then give a very good opportunity to falsify the presented model. If at either of these flybys an anomaly would be detected, the model is then definitely falsified, regarding the investigated equations.

## 7.2 Spatially-fixed software filter for Earth gravity field determination

It has been pointed out above, why the assumed field term has not been discovered up to now, if it exists. One major point is the lack of a software filter with fixed spatial orientation, or the lack of motivation for the search of such a component, respectively. Now the more precisely formulated field term could be used to develop an appropriate filter or to effectively use an existing filter for screening the old orbit data. This would be a further quick and cost-efficient way to falsify or to confirm the proposed model.

## 7.3 Reprocessing of old flyby data by adapted orbit determination software

Another in principle simple test would be the investigation, whether, with the assumption of the presented field term, the discontinuities in the measured Doppler residuals for the Galileo1 and the NEAR Earth flyby could be removed just as good or even better than with the formerly estimated hypothetical Earth gravity field described in [2]. To this purpose, access to better data would be necessary. Potentially, it would be sufficient to reprocess the old data by adaption of equation (1) to the orbit determination software used to calculate the Doppler residuals.

## 7.4  Mission with drag-free control system at low perigee altitude

Lämmerzahl, Preuss and Dittus [4] have proposed such a dedicated flyby mission in case of a measured anomaly at the Rosetta mars flyby. This proposal has been greatly supported by the results of the present investigation, especially by the sharp altitude dependency. According to these results optimal orbit parameters could possibly be chosen to verify or to falsify the predictions.

## 7.5  Independent confirmation of the presented simulation results

The results of the simulation presented here are based on a relatively complex program structure. Considerable effort has been made to find and to eliminate programming errors. The correctness of the results has been verified in a plurality of special tests. Nevertheless, it cannot be ruled out, that the achieved results are based on partly incorrect calculation. Considering the importance of the results in the case of correctness, it would be desirable to have the presented simulation reproduced by an independent party.

# 8  Conclusions

The observed flyby anomaly values of the spacecrafts Galileo, NEAR, Rosetta, Cassini and Messenger could be simulated by assumption of a gravitational field after equations (1) and (2). According to the simulation, this field has an asymmetry, determined by the motion of Earth against a reference frame, not coinciding with the CMB reference frame. The orientation of this reference frame to the motion of the Sun and the values of the arbitrary parameters $A$, $B$ and $C$ of equation (1) could be determined in such a way, that for the flybys Galileo1, NEAR, Rosetta1 and Cassini the nominal values of the measured anomaly were calculated, while for Galileo2 and Messenger plausible results were obtained, compliant with the uncertain measuring data.
Furthermore, there was qualitative congruence with the curve of the measured Doppler residuals immediately after the perigee of Galileo1, and as to be expected similar acceleration curves with respect to formerly estimated hypothetical gravity accelerations for the NEAR flyby.
The remarkable compliance of the simulation results with the six observed anomaly values und additionally with the observed Doppler residual curve as shown in Table 4 and Figure 3 was obtained by variation of only five arbitrary parameters. This seems to be a clear indication of an underlying physical reality in contrast to an only numerical effect.
With the same parameter set as for the Earth, the simulation was carried out for the flybys of Rosetta at Mars on 25.02.07 and at Earth on 13.11.07 and 13.11.09, whereas the viability of the Earth-derived parameters $A$, $B$ and $C$ for the Mars flyby cannot be justified. The simulation leads to the prediction, that for the Mars flyby of Rosetta, with great uncertainty, an anomaly of 5mm/s is to be expected, while for the following Earth flybys definitely no measurable anomaly will be observed.
The assumed asymmetric deformation of the gravitational field in direction of the Earth's motion against a unique reference frame violates the Principle of Relativity. Therefore, the proposed equation (1) cannot be derived by current theory. The structure of the additional field term suggests a relationship to transport phenomena, implying the possibility of a gravitational component without coupling to matter. In contrast to gravitational bound baryonic matter condensations, such a component would take part on the cosmic expansion, and would have properties normally attributed to the so-called Dark Matter.
In case of the reality of the proposed or a similar gravitational field, the challenge arises to develop a new theory, which correctly describes the new phenomena, while including General Relativity as a limit case. A rudimental approach justified by numerical calculation is given above.

## Acknowledgement

I thank my wife Dorothea for generously tolerating and encouraging my investigations on this subject over more than 6 months, Dr. Alexander Unzicker for valuable hints, Sara Meitner for reading over the manuscript, and C. Lämmerzahl, O. Preuss and H. Dittus for publishing a short version of their paper [4] in the journal 'Sterne und Weltraum', issue 4/2007, where the flyby anomaly came to my attention.